\documentclass[a4paper,10pt]{article}
\usepackage[dvips]{graphicx}
\usepackage{amssymb,amsmath}
\numberwithin{equation}{section}

\begin{document}

 \def\gsim{ \lower .75ex \hbox{$\sim$} \llap{\raise .27ex \hbox{$>$}} }
 \def\lsim{ \lower .75ex \hbox{$\sim$} \llap{\raise .27ex \hbox{$<$}} }

% \begin{titlepage}

% \begin{flushright}
% arXiv:2012:5266
% \end{flushright}

 \title{\Large \bf FRW   Cosmology in  $F(R,T)$ Gravity}

\author{Ratbay  Myrzakulov\footnote{Email: rmyrzakulov@gmail.com; rmyrzakulov@csufresno.edu}
 \\ \textit{Eurasian International Center for Theoretical Physics and  Department of General } \\ \textit{ $\&$  Theoretical Physics, Eurasian National University, Astana 010008, Kazakhstan}}

 %\pacs{04.50.-h, 04.20.Fy, 98.80.-k, 95.36.+x}
\date{}
 \maketitle

% \end{titlepage}

 \renewcommand{\baselinestretch}{1.1}

 \begin{abstract}
In this paper, we consider a theory of gravity with a
metric-dependent torsion namely the  $F(R,T)$ gravity, where $R$ is
the curvature scalar and $T$ is the torsion scalar. We study the
geometric root of such theory. In particular we give the derivation
of the model from the geometrical point of view. Then we present the
more general form of $F(R,T)$ gravity with two arbitrary functions
and give some of its particular cases. In particular, the usual
$F(R)$ and $F(T)$ gravity theories are particular cases of the
$F(R,T)$ gravity. In the cosmological context, we find that our new
gravitational theory can describe the accelerated expansion of the
Universe.
 \end{abstract}

 \tableofcontents
\section{Introduction}

The discovery of the accelerated expansion of the Universe has
revolutionized  modern  cosmology. It is generally assumed that this
cosmic acceleration is due to some kind of energy-matter with
negative pressure known as `dark energy' (DE). The nature of DE as
well as its cosmological origin remain elusive so far. In order to
explain the nature of the DE and the accelerated expansion, a
variety of theoretical models have been proposed in the literature,
such as quintessence \cite{j1}, phantom energy \cite{j2}, k-essence
\cite{j3}, tachyon \cite{j4}, f-essence \cite{j5}, Chaplygin gas
\cite{j6}, g-essence, etc. Among these different models of DE, the
modified gravity models are quite interesting as they incorporate
some notions of the quantum and general (classical) gravity
theories. There are several modified gravity theories like $F(R)$
gravity, $F(G)$ gravity, $F(T)$ gravity and so on (see e.g.
Refs.\cite{r1}-\cite{Ba}). In our opinion, one of interesting and
prospective versions of modified gravity theories is the $F(R,T)$
gravity, where $R$ is the curvature scalar and $T$ is the torsion
scalar. It can reproduce the unification of $F(R)$ and $F(T)$
gravity theories.  Recently one of the versions of $F(R,T)$ gravity
was proposed in \cite{MK1} and its some properties was studied in
\cite{MK1A}.  In this paper we continue our work on $F(R,T)$
gravity.

In the  previous papers \cite{MK1}-\cite{MK1A}, we introduced our $F(R,T)$
gravity models in an "ad hoc" manner. Contrary to previous, here we
demonstrate (in the example of the M$_{37}$ - model) that $F(R,T)$
gravity models can be derived from geometry. So we put $F(R,T)$
gravity models in the geometrical language which is traditionally
done for the modern gravity theories \cite{T}-\cite{r5}. Also we show
that $F(R,T)$ gravity models can describe the accelerated expansion
of the Universe including the phantom crossing case.  Finally  we
find a cosmological solution of the $F(R,T)$ gravity  model
corresponding to  the de Sitter Universe.

As is well-known, General Relativity (GR) is described by Riemannian
geometry which is torsion-free. In literature, several gravitational
theories with torsion were proposed (see e.g. Refs. \cite{C1}-\cite{Poplawski})
which show that the torsion effects should be included in the
extensions of GR. In these theories, usually torsion is not
propagating, since it is given algebraically in terms of the spin
matter fields. As a consequence, the torsion can only be detected in
the presence of spin matter fields. In contrast, in $F(T)$ gravity
(here $T$ is the torsion scalar)  torsion has no source  and it can
propagate without any matter source. This property of $F(T)$ gravity
is same as the corresponding property of $F(R)$ gravity where the
curvature ($R$) can  propagates without the source. In  our $F(R,T)$
gravity, at least  at cosmological level it means that in a flat FRW
space-time, the torsion can propagate in the absence of spin matter
fields as in $F(T)$ gravity. In other words, our $F(R,T)$ gravity
inherits the relevant properties of its constituent two theories -
$F(R)$ and $F(T)$ gravity theories. In other words, in our $F(R,T)$
gravity torsion as well as curvature can propagate without any
matter source. It is a main merit (advantage) of our model.  In fact
this is a crucial point, otherwise the additional scalar torsion
degree of freedom are not different from the additional metric
gravitational  degree of freedom present in  extended $F(R)$ models.
In other words, torsion $T$ is a fundamental quantity like curvature
$R$.  The  $F(R,T)$ theory is considered as a fundamental
gravitational theory, which describes the evolution of our Universe.
Another important advantage of our $F(R,T)$ gravity models is that
they contain as particular cases - two well-known modified gravity
theories namely $F(R)$ gravity and $F(T)$ gravity. It indicates that
our $F(R,T)$ gravity models are direct generalizations of
well-studied and well-known $F(R)$ and $F(T)$ gravity theories and
is the unified description of them.

A few words about our  notations. First we  would like to draw the
attention to the existence to two types of gravity theory with the
same name;  that is, there are two type of $F(R,T)$ gravity
theories. One was proposed in \cite{Harko} but in this case, $T$ is
the trace of the energy-moment tensor. Second type is proposed in
\cite{MK1} (see also \cite{MK1A}) and also called $F(R,T)$ gravity and
which we will study in this work. Lastly, we note that in some of
our papers including this one, we used some conventional notations
to distinguish the different gravitational (cosmological) models
that were proposed (see \cite{MK1}). For example in this paper we use
notations like the M$_{43}$-model, the M$_{37}$-model and so on.
These notations do not bring any physical or mathematical meaning
and used just to distinguish and fix different models for our
convenience.

This paper is organized as follows: In section-2, the gravitational
action of $F(R,T)$ gravity and its arguments are derived from the
geometrical point of view. Using them to the spatially flat FRW
metric, an action with the curvature and torsion scalars is
obtained. The Lagrangian formulation of the generalized $F(R,T)$
gravity model is given in section-3. Section-4 is devoted to
construct some cosmological solutions for a particular model: $F=\mu
R+\nu T$. In this case, the solutions of the cosmological equations
are divided into two classes. Each of them is related with some
torsion scalar functions. In particular, the exact de Sitter
solution is found. These exact analytic solutions of the
cosmological equations describe the accelerated expansion of the
Universe. Section 5 is devoted to the conclusion.

\section{Geometrical roots of $F(R,T)$ gravity}

We start from the M$_{43}$ - model (about our   notations,  see e.g.
Refs.  \cite{MK1}-\cite{MK1A}). This model is one of the
representatives of $F(R,T)$ gravity.
 The  action of the M$_{43}$ - model reads as
\begin{eqnarray}
 S_{43}&=&\int d^4 x\sqrt{-g}[F(R,T)+L_m],\nonumber\\
 R&=&R_s=\epsilon_1g^{\mu\nu}R_{\mu\nu},\label{2.1}\\
   T&=&T_s=\epsilon_2{S_\rho}^{\mu\nu}\,{T^\rho}_{\mu\nu},\nonumber
 \end{eqnarray} where $L_m$ is the matter Lagrangian, $\epsilon_i=\pm 1$ (signature), $R$ is the curvature scalar, $T$ is the torsion scalar (about our notation see below). In this section we try to
  give one of the possible geometric formulations of  M$_{43}$ - model. Note that we have different cases related with the signature:
 (1) $\epsilon_1=1,  \epsilon_2=1$; (2) $\epsilon_1=1,  \epsilon_2=-1$;
 (3) $\epsilon_1=-1,  \epsilon_2=1$;  (4) $\epsilon_1=-1,  \epsilon_2=-1$.
  Also note that M$_{43}$ - model is a particular case of M$_{37}$ - model having the form
  \begin{eqnarray}
  S_{37}&=&\int d^4 x\sqrt{-g}[F(R,T)+L_m],\nonumber\\
 R&=&u+R_s=u+\epsilon_1g^{\mu\nu}R_{\mu\nu},\label{2.2}\\
   T&=&v+T_s=v+\epsilon_2{S_\rho}^{\mu\nu}\,{T^\rho}_{\mu\nu},\nonumber
 \end{eqnarray}
 where
 \begin{equation}\label{2.3}
R_s=\epsilon_1g^{\mu\nu}R_{\mu\nu},\quad T_s=\epsilon_2{S_\rho}^{\mu\nu}\,{T^\rho}_{\mu\nu}
 \end{equation}
 are the standard forms of the curvature and torsion scalars.

\subsection{General case}

To understand the geometry of the M$_{43}$ - model,   we consider
some   spacetime with the curvature and torsion so that its
connection  $G^\lambda {}_{\mu \nu }$ is a sum of  the curvature and
torsion parts.
 In this paper,  the Greek alphabets ($\mu $, $\nu $%
, $\rho $, $...=0,1,2,3$)  are related to spacetime, and the Latin
alphabets ($i,j,k, ...=0,1,2,3$)  denote
 indices, which are raised and lowered with
the Minkowski metric $\eta _{ij}$ $=$ diag ($-1,+1,+1,+1$).
  For our spacetime  the  connection  $G^\lambda {}_{\mu \nu }$ has the form
\begin{equation}\label{2.4}
G^\lambda {}_{\mu \nu }=e_i{}^\lambda \partial _\mu
e^i{}_\nu
+e_j{}^\lambda e^i{}_\nu \omega{}^j{}_{i\mu }=\Gamma^\lambda {}_{\mu \nu }+K^\lambda {}_{\mu \nu}.
 \end{equation}
Here $\Gamma^{j}_{i\mu}$ is the Levi-Civita connection and $K^{j}_{i\mu}$ is the contorsion.
 Let the metric has the form
\begin{equation}\label{2.5}
 ds^2=g_{ij}dx^idx^j.
 \end{equation}
  Then the  orthonormal tetrad
 components $e_i(x^\mu)$ are related to the metric through
 \begin{equation}\label{2.6}
 g_{\mu\nu}=\eta_{ij}e_\mu^i e_\nu^j,
 \end{equation}
so that the orthonormal condition reads as
\begin{equation}\label{2.7}
 \eta_{ij}=g_{\mu\nu}e^\mu_i e^\nu_j.
 \end{equation}
 Here $\eta_{ij}=diag(-1,1,1,1)$, where we used the relation
 \begin{equation}\label{2.8}
 e_\mu^i e^\mu_j=\delta^i_j.
 \end{equation}
 The quantities $\Gamma^{j}_{i\mu}$ and  $K^{j}_{i\mu}$ are defined as
 \begin{equation}\label{2.9}
\Gamma^l{}_{jk}=\frac{1}{2}g^{lr} \{\partial _k g_{rj} + \partial _j
g_{rk} - \partial _r g_{jk} \}
\end{equation}
 and
 \begin{equation}\label{2.10}
K^{\lambda}_{\mu\nu}=-\frac 12\left( T^\lambda {}_{\mu \nu
}+T_{\mu
\nu }{}^\lambda +T_{\nu \mu }{}^\lambda \right),
\end{equation}
respectively.
Here the components of the torsion tensor are given by
 \begin{eqnarray}\label{2.11}
T^\lambda {}_{\mu \nu } &=&e_i{}^\lambda T^i{}_{\mu \nu }=G^\lambda
{}_{\mu \nu }-G^\lambda {}_{\nu \mu },   \\ \label{2.12}
T{}^i{}_{\mu \nu } &=&\partial _\mu e{}^i{}_\nu -\partial _\nu
e{}^i{}_\mu +G{}^i{}_{j\mu }e{}^j{}_\nu -G{}^i{}_{j\nu
}e{}^j{}_\mu.
\end{eqnarray}

The curvature $R^\rho {}_{\sigma \mu \nu }$ is defined  as
\begin{eqnarray} \label{2.13}
R^\rho {}_{\sigma \mu \nu } &=&e_i{}^\rho e^j{}_\sigma R^i{}_{j\mu
\nu }=\partial _\mu G^\rho {}_{\sigma \nu }-\partial _\nu
G^\rho {}_{\sigma \mu }+G^\rho {}_{\lambda \mu }G^\lambda {}_{\sigma \nu }-G^\rho {}_{\lambda \nu }G^\lambda {}_{\sigma \mu }
\nonumber \\
&=&\bar{R}^\rho {}_{\sigma \mu \nu }+\partial _\mu
K^\rho {}_{\sigma \nu }-\partial _\nu K^\rho {}_{\sigma \mu }+K^\rho
{}_{\lambda \mu }K^\lambda {}_{\sigma \nu }-K^\rho {}_{\lambda \nu
}K^\lambda {}_{\sigma
\mu }  \nonumber \\
&&+\Gamma^{\rho}_{\lambda\mu} K^\lambda {}_{\sigma \nu
}-\Gamma^{\rho}_{\lambda\nu} K^\lambda {}_{\sigma \mu
}+\Gamma^{\lambda}_{\sigma\nu} K^\rho {}_{\lambda \mu
}-\Gamma^{\lambda}_{\sigma\mu} K^\rho {}_{\lambda \nu
},
\end{eqnarray}
where  the Riemann curvature of the Levi-Civita connection is
defined in the standard way
\begin{equation} \label{2.14}
\bar{R}^\rho {}_{\sigma \mu \nu }=\partial
_\mu \Gamma^{\rho}_{\sigma\nu} -\partial _\nu \Gamma^{\rho}_{\sigma\mu} +\Gamma^{\rho}_{\lambda\mu}
\Gamma^{\lambda}_{\sigma\nu} -\Gamma^{\rho}_{\lambda\nu} \Gamma^{\lambda}_{\sigma\mu}. \end{equation}
 Now we introduce two important quantities namely the curvature $(R)$ and torsion $(T)$ scalars as
\begin{eqnarray} \label{2.15}
R&=&g^{ij}R_{ij}, \\   \label{2.16}
T&=&S_\rho^{\mu\nu}T^\rho_{\mu\nu},
\end{eqnarray}
where
\begin{equation} \label{2.17}
        S_\rho {}^{\mu\nu}=\frac{1}{2}\left(K_\rho{}^{\mu\nu}+\delta_{\rho}^\mu T_\theta{}^{\theta\nu}-\delta_{\rho}^\nu T_\theta{}^{\theta\mu}\right).
\end{equation}
Then the M$_{43}$ - model we write in the form \eqref{2.1}. To conclude this subsection, we note that in GR, it is postulated that $T^\lambda {}_{\mu \nu }=0$ and such 4-dimensional spacetime  manifolds with metric and without torsion are labelled as V$_{4}$. At the same time, it is a general convention to call U$_{4}$,  the manifolds endowed with metric and torsion.

\subsection{FRW  case}

From here we work with  the spatially flat FRW metric
\begin{equation}\label{2.18}
 ds^2=-dt^2+a^2(t)(dx^2+dy^2+dz^2),
 \end{equation}where $a(t)$ is the scale factor. In this case,
 the non-vanishing components of the Levi-Civita connection are
\begin{eqnarray}\label{2.19}
\Gamma^{0}_{00}&=&\Gamma^{0}_{0i}=\Gamma^{0}_{i0}=\Gamma^{i}_{00}=\Gamma^{i}_{jk}=0,\nonumber \\
\Gamma^{0}_{ij}&=& a^2H
\delta _{ij},  \\
\Gamma^{i}_{jo}&=&\Gamma^{i}_{0j}=H\delta _j^i,\nonumber
\end{eqnarray}
where $H=(\ln a)_t$ and $i,j,k,...=1,2,3.$ Now let us calculate the
components of torsion tensor. Its non-vanishing  components  are
given by:
\begin{eqnarray} \label{2.20}
T_{110} &=&T_{220}=T_{330}=a^2h, \nonumber  \\
T_{123} &=&T_{231}=T_{312}=2a^3f,
\end{eqnarray}
where  $h$ and  $f$ are some real functions (see e.g. Refs. \cite{T}).
 Note that  the indices of the torsion tensor are raised and lowered with respect to the metric, that is
\begin{equation}\label{2.21}
 T_{ijk}=g_{kl}T_{ij}{}{}^{l}.
 \end{equation}

Now we can find  the contortion components. We get
\begin{eqnarray} \label{2.22}
K^1{}_{10} &=&K^2{}_{20}=K^3{}_{30}=0,  \nonumber \\
K^1{}_{01} &=&K^2{}_{02}=K^3{}_{03}=h,  \nonumber \\
K^0{}_{11} &=&K^0{}_{22}=K^0{}_{22}={}a^2h, \\
K^1{}_{23} &=&K^2{}_{31}=K^3{}_{12}=-af,  \nonumber \\
K^1{}_{32} &=&K^2{}_{13}=K^3{}_{21}=af. \nonumber
\end{eqnarray}

The non-vanishing components of the curvature $R^\rho {}_{\sigma \mu
\nu }$  are given by
\begin{eqnarray} \label{2.23}
R^0{}_{101} &=&R^0{}_{202}=R^0{}_{303}=a^2( \dot{H}%
+H^2+Hh+\dot{h}),  \nonumber \\
R^0{}_{123} &=&-R^0{}_{213}=R^0{}_{312}=2a^3f( H+h),
\nonumber
\\
R^1{}_{203} &=&-R^1{}_{302}=R^2{}_{301}=-a( Hf+\dot{f}
),  \nonumber \\
R^1{}_{212} &=&R^1{}_{313}=R^2{}_{323}=a^2[ ( H+h)
^2-f^2].
\end{eqnarray}
Similarly, we write the non-vanishing components of the  Ricci
curvature tensor $R{}_{\mu \nu }$ as
\begin{eqnarray} \label{2.24}
R{}_{00} &=&-3\dot{H}-3\dot{h}-3H^2-3Hh,
\nonumber
\\
R{}_{11} &=&R{}_{22}=R{}_{33}=a^2( \dot{H}+\dot{h}+3H^2+5Hh+2h^2-f^2).
\end{eqnarray}
At the same time, the non-vanishing components of the tensor $S_\rho^{\mu\nu}$ are given by
\begin{eqnarray} \label{2.25}
S_1^{10}&=&\frac{1}{2}\left(K^{10}_1+\delta_1^1T^{\theta 0}_\theta-\delta_1^0T^{\theta\nu}_\theta\right)=\frac{1}{2}\left(h+2h\right)=h,\\ \label{2.26}
S_1^{10}&=&S_2^{20}=S_3^{30}=2h,\\ \label{2.27}
S_1^{23}&=&\frac{1}{2}\left(K^{23}_1+\delta_1^2+\delta_1^3\right)=-\frac{f}{2a},\\ \label{2.28}
S_1^{23}&=&S_2^{31}=S_3^{21}=-\frac{f}{2a}
\end{eqnarray}
and
\begin{eqnarray}\label{2.29}
T=T^1_{10}S_1^{10}+T^2_{20}S_2^{20}+T^3_{30}S_3^{30}+T_1^{23}S^1_{23}+T^2_{31}S_2^{31}+T^3_{12}S_3^{12}.
\end{eqnarray}
Now we are ready to write the explicit forms of the curvature and torsion scalars. We have
\begin{eqnarray} \label{2.30}
R&=&6(\dot{H}+2H^2) +6\dot{h}+18Hh+6h^2-3f^2 \\ \label{2.31}
T&=&6(h^2-f^2).
\end{eqnarray}
So finally for the FRW metric, the M$_{43}$ - model  takes the form
\begin{eqnarray}
S_{43}&=&\int d^4 x\sqrt{-g}[F(R,T)+L_m], \nonumber\\
R&=&6(\dot{H}+2H^2) +6\dot{h}+18Hh+6h^2-3f^2, \label{2.32}\\
T&=&6(h^2-f^2).\nonumber
\end{eqnarray}
It (that is the M$_{43}$ - model)  is one of geometrical realizations of $F(R,T)$ gravity in the sense that it was derived from the purely geometrical point of view.
\subsection{Particular cases}
The M$_{43}$ - model admits some important features from the
physical and
 geometrical point of view. In this subsection we want to
  present some particular reductions of the M$_{43}$ - model \eqref{2.32} such as the FRW metric case.

\subsubsection{$F(R)$ gravity}

Let $h=f=0$. Then the M$_{43}$ - model \eqref{2.32} reduces to the
case
\begin{eqnarray}
S_{43}&=&\int d^4 x\sqrt{-g}[F(R_s)+L_m], \nonumber\\
R&=&R_s=6(\dot{H}+2H^2), \label{2.33}\\
T&=&0,\nonumber
\end{eqnarray}
which is the usual $F(R)$ gravity.

\subsubsection{The M$_{43A}$ - model} Let $f=0$. Then the M$_{43}$ - model \eqref{2.32} reduces to the case
\begin{eqnarray}
S_{43A}&=&\int d^4 x\sqrt{-g}[F(R,T)+L_m], \nonumber\\
R&=&R_s=6(\dot{H}+2H^2) +6\dot{h}+18Hh+6h^2, \label{2.34}\\
T&=&6h^2.\nonumber
\end{eqnarray}
We can note that for this particular case $T\geq 0$.
\subsubsection{The M$_{43B}$ - model} Let $h=0$. Then we get the M$_{43B}$ - model  with the action
\begin{eqnarray}
S_{43B}&=&\int d^4 x\sqrt{-g}[F(R,T)+L_m], \nonumber\\
R&=&6(\dot{H}+2H^2) -3f^2, \label{2.35}\\
T&=&-6f^2.\nonumber\end{eqnarray}
For this particular model $T\leq 0$ as follows from the last equation.

\subsubsection{The M$_{43C}$ - model} Let $f=\pm h$.
Then $T=0$ and obtain the following M$_{43C}$ - model:
\begin{eqnarray}
S_{43C}&=&\int d^4 x\sqrt{-g}[F(R)+L_m],\label{2.36} \\
R&=&6(\dot{H}+2H^2) +6\dot{h}+18Hh+3h^2. \label{2.37}
\end{eqnarray}
Hence as $h=0$ we get the usual $F(R)$  gravity with the scalar curvature $R=R_s=6(\dot{H}+2H^2)$.

\subsubsection{The M$_{43D}$ - model}
 Let $f=\pm\sqrt{2(\dot{H}+2H^2) +2\dot{h}+6Hh+2h^2}$. Then $R=0$ and we have the M$_{43D}$ - model:
\begin{eqnarray}
S_{43D}&=&\int d^4 x\sqrt{-g}[F(T)+L_m], \label{2.38}\\
T&=&-12(\dot{H}+2H^2+\dot{h}+3Hh+0.5h^2).\label{2.39}\end{eqnarray}

\subsubsection{The M$_{43E}$ - model}
 Now we assume that $h=0$ and  $f=\pm\sqrt{2(\dot{H}+2H^2)}$. Then $R=0$ and we have the M$_{43D}$ - model:
\begin{eqnarray}
S_{43E}&=&\int d^4 x\sqrt{-g}[F(T)+L_m], \label{2.40}\\
T&=&-12(\dot{H}+2H^2).\label{2.41}\end{eqnarray}

\section{Lagrangian formulation of $F(R,T)$ gravity}

Of course, we can work with the form \eqref{2.32} of $F(R,T)$
gravity. But a more interesting  and general form of $F(R,T)$ gravity is the
so-called M$_{37}$ -  model. The action of the M$_{37}$ -
gravity reads as  \cite{MK1}
\begin{eqnarray}
S_{37}&=&\int d^4 x\sqrt{-g}[F(R,T)+L_m],\nonumber \\
 R&=&u+R_s=u+6\epsilon_1(\dot{H}+2H^2),\label{3.1}\\
   T&=&v+T_s=v+6\epsilon_2H^2,\nonumber
 \end{eqnarray}
 where
 \begin{equation}\label{3.2}
 R_s=6\epsilon_1(\dot{H}+2H^2),\quad T_s=6\epsilon_2H^2.
 \end{equation}
So we can see that here instead of two functions $h$ and $f$ in
\eqref{2.32}, we introduced two new functions $u$ and $v$. For
example, for \eqref{2.32} these functions have the form
\begin{eqnarray}
 u&=&6(1-\epsilon_1)(\dot{H}+2H^2)+ 6\dot{h}+18Hh+6h^2-3f^2,\label{3.3} \\
  v&=&6(h^2-f^2-\epsilon_2H^2)\label{3.4}
 \end{eqnarray}
 that again tells us that the M$_{43}$ - model is a  particular case of  M$_{37}$ - model [Note that if $\epsilon_1=1=\epsilon_2$ we have $u= 6\dot{h}+18Hh+6h^2-3f^2,\quad v=6(h^2-f^2-H^2)$].
But in general we think (or assume) that   $u=u(t, a,\dot{a}, \ddot{a},\dddot{a}, ...; f_i)$
 and $v=v(t,a,\dot{a}, \ddot{a},\dddot{a}, ...; g_i)$,   while $f_i$ and $g_i$ are some unknown
  functions related with the geometry of the spacetime. So below we will work
  with a
  more general form of $F(R,T)$ gravity namely the  M$_{37}$ - gravity  \eqref{3.1}.
 Introducing the Lagrangian multipliers we now can rewrite the action (\ref{3.1}) as
 \begin{equation}\label{3.5}
 S_{37}=\int dt\,a^3\left\{F(R,T)-\lambda\left[T-v-
 6\epsilon_2\frac{\dot{a}^2}{a^2}\right]-\gamma\left[R-u- 6\epsilon_1\left(\frac{\ddot{a}}{a}+\frac{\dot{a}^2}{a^2}\right)\right]+L_m\right\},
 \end{equation}
 where $\lambda$ and $\gamma$ are Lagrange multipliers. If we take the variations with
 respect to $T$ and $R$ of this action, we get
  \begin{equation}\label{3.6}
 \lambda=F_T, \quad \gamma=F_R.
 \end{equation}
 Therefore, the action~(\ref{3.5}) can be rewritten as
 \begin{equation}\label{3.7}
 S_{37}=\int dt\,a^3\left\{F(R,T)-F_T\left[T-v-
 6\epsilon_2\frac{\dot{a}^2}{a^2}\right]-F_R\left[R-u- 6\epsilon_1
 \left(\frac{\ddot{a}}{a}+\frac{\dot{a}^2}{a^2}\right)\right]+L_m\right\}.
\end{equation}
 Then the  corresponding  point-like Lagrangian reads
 \begin{equation} \label{3.8}
 L_{37}=a^3[F-(T-v)F_T-(R-u)F_R+L_m]-
 6(\epsilon_1F_R-\epsilon_2F_T) a\dot{a}^2-6\epsilon_1(F_{RR}\dot{R}+F_{RT}\dot{T})a^2\dot{a}.
\end{equation}
As is well known, for our dynamical system, the Euler-Lagrange
 equations read as
 \begin{eqnarray}
 \frac{d}{dt}\left(\frac{\partial  L_{37}}{\partial \dot{R}}
 \right)-\frac{\partial  L_{37}}{\partial R}&=&0,\label{3.9}\\
  \frac{d}{dt}\left(\frac{\partial  L_{37}}{\partial\dot{T}}
 \right)-\frac{\partial  L_{37}}{\partial T}&=&0,\label{3.10}\\
 \frac{d}{dt}\left(\frac{\partial  L_{37}}{\partial\dot{a}}
 \right)-\frac{\partial  L_{37}}{\partial a}&=&0.\label{3.11}
 \end{eqnarray}
   Hence, using the expressions
 \begin{eqnarray}
 \frac{\partial  L_{37}}{\partial \dot{R}}&=&-6\epsilon_1F_{RR}a^2\dot{a}, \label{3.12}\\
 \frac{\partial  L_{37}}{\partial \dot{T}}&=&-6\epsilon_1F_{RT}a^2\dot{a},\label{3.13}\\
 \frac{\partial  L_{37}}{\partial \dot{a}}&=&-12(\epsilon_1F_R-\epsilon_2F_T)a \dot{a}-6\epsilon_1(F_{RR}\dot{R}+F_{RT}\dot{T}+F_{R\psi}\dot{\psi})a^2+a^3F_Tv_{\dot{a}}+a^3F_Ru_{\dot{a}},\quad \label{3.14}
 \end{eqnarray}
 we get
\begin{eqnarray}
 a^3 F_{TT}\left(T-v-
 6\epsilon_2\frac{\dot{a}^2}{a^2}\right)&=&0,\\ \label{eq15}
 a^3 F_{RR}\left(R-u-
 6\epsilon_1(\frac{\ddot{a}}{a}+\frac{\dot{a}^2}{a^2})\right)&=&0,\\ \label{eq16}
U+B_2F_{TT}+B_1F_{T}+C_2F_{RRT}+C_1F_{RTT}+C_0F_{RT}+MF+6\epsilon_2a^2p&=&0,\label{eq17}
 \end{eqnarray}
 respectively. Here
 \begin{eqnarray}
 U&=&A_3F_{RRR}+A_2F_{RR}+A_1F_{R},\\
 A_3&=&-6\epsilon_1\dot{R}^2a^2,\label{eq20}\\
 A_2&=&-12\epsilon_1\dot{R}a\dot{a}-6\epsilon_1\ddot{R}a^2+a^3\dot{R}u_{\dot{a}},\\
   A_1&=&12\epsilon_1\dot{a}^2+6\epsilon_1a \ddot{a}+3a^2\dot{a}u_{\dot{a}}+a^3\dot{u}_{\dot{a}}-a^3u_a,\\
 B_2&=&12\epsilon_2\dot{T}a \dot{a}+a^3\dot{T}v_{\dot{a}},\\
 B_1&=&24\epsilon_2\dot{a}^2+12\epsilon_2a \ddot{a}+3a^2\dot{a}v_{\dot{a}}+a^3\dot{v}_{\dot{a}}-a^3v_a,\\
 C_2&=&-12\epsilon_1a^2\dot{R}\dot{T},\\
 C_1&=&-6\epsilon_1a^2\dot{T}^2,\\ C_0&=&-12\epsilon_1\dot{T}a\dot{a}+12\epsilon_2\dot{R}a\dot{a}-6\epsilon_1a^2\ddot{T}+a^3\dot{R}v_{\dot{a}}+a^3\dot{T}u_{\dot{a}},\\
 M&=&-3a^2.
 \end{eqnarray}
 If $F_{RR}\neq 0, \quad F_{TT}\neq 0$, from Eqs.~(\ref{eq16}) and (\ref{eq17}), it is easy to find that
\begin{equation}
 R=u+6\epsilon_1(\dot{H}+2H^2),\quad   T=v+6\epsilon_2H^2,
 \end{equation}
 so that  the relations ~(\ref{3.1}) are  recovered. Generally, these
 equations are  the Euler constraints of the dynamics. Here  $a, R, T$ are the generalized coordinates of
 the configuration space. On the other hand, it is also
 well known that the total energy (Hamiltonian) corresponding
 to Lagrangian $ L_{37}$ is given by
\begin{equation}\label{eq30}
  H_{37}=\frac{\partial  L_{37}}
 {\partial\dot{a}}\dot{a}+\frac{\partial  L_{37}}
 {\partial\dot{R}}\dot{R}+\frac{\partial  L_{37}}
 {\partial\dot{T}}\dot{T}-{ L_{37}}.
 \end{equation}
 Hence using (\ref{3.12})-(\ref{3.14}) we obtain
  $$
  H_{37}=[-12(\epsilon_1F_R-\epsilon_2F_T)a \dot{a}-6\epsilon_1(F_{RR}\dot{R}+F_{RT}\dot{T}+F_{R\psi}\dot{\psi})a^2+a^3F_Tv_{\dot{a}}+a^3F_Ru_{\dot{a}}]\dot{a}$$
  $$
-6\epsilon_1F_{RR}a^2\dot{a}\dot{R}-6\epsilon_1F_{RT}a^2\dot{a}\dot{T}-[a^3(F-TF_T-RF_R+vF_T+uF_R+L_m)-
 $$
  \begin{equation}6(\epsilon_1F_R-\epsilon_2F_T) a\dot{a}^2
-6\epsilon_1(F_{RR}\dot{R}+F_{RT}\dot{T}+F_{R\psi}\dot{\psi})a^2\dot{a}].
 \end{equation}
 Let us rewrite this formula as
 \begin{equation}\label{eq31}
 H_{37}=D_2F_{RR}+D_1F_R+JF_{RT}+E_1F_T+KF+2a^3\rho,
 \end{equation}
 where
 \begin{eqnarray}
 D_2&=&-6\epsilon_1\dot{R}a^2\dot{a},\\
   D_1&=&6\epsilon_1a \ddot{a}+a^3u_{\dot{a}}\dot{a},\\
  J&=&-6\epsilon_1a^2 \dot{a}\dot{T},\\
 E_1&=&12\epsilon_2a \dot{a}^2+a^3v_{\dot{a}}\dot{a},\\
 K&=&-a^3.
 \end{eqnarray}
As usual we assume  that the total energy $H_{37}=0$ (Hamiltonian constraint). 
 So finally we have the following equations of the M$_{37}$ -  model \cite{MK1}-\cite{MK1A}:
\begin{eqnarray} \label{3.9}
 D_2F_{RR}+D_1F_R+JF_{RT}+E_1F_T+KF&=&-2a^3\rho,\nonumber\\
   U+B_2F_{TT}+B_1F_{T}+C_2F_{RRT}+C_1F_{RTT}+C_0F_{RT}+MF  &=&6a^2p,\label{1.12}\\
 \dot{\rho}+3H(\rho+p)&=&0.\nonumber
 \end{eqnarray}

 It deserves to note that the M$_{37}$ - model (\ref{3.1}) admits some interesting particular
  and physically important  cases. Some particular cases are now presented.

 \textit{i) The M$_{44}$ - model.}
 Let the function $F(R,T)$ be independent from the torsion scalar $T$: $F=F(R,T)=F(R)$.
  Then the  action (\ref{3.1}) acquires the form
   \begin{equation}\label{3.25}
 S_{44}=\int d^4 x e[F(R)+L_m],
 \end{equation}
 where
 \begin{equation}\label{3.26}
 R=u+R_s=u+\epsilon_1g^{\mu\nu}R_{\mu\nu},
 \end{equation}
  is the curvature scalar. It is the M$_{44}$ - model. We work with  the FRW metric.
   In this case  $R$ takes  the form
 \begin{equation} \label{3.27}
R=u+6\epsilon_1(\dot{H}+2H^2).
\end{equation}
The action can be rewritten as
 \begin{equation}\label{3.28}
 S_{44}=\int dtL_{44},
 \end{equation}
 where the   Lagrangian is given by
 \begin{equation}\label{3.29}
  L_{44}=a^3[F-(R-u)F_R+L_m]-
 6\epsilon_1F_Ra\dot{a}^2-6\epsilon_1F_{RR}\dot{R}a^2\dot{a}.
 \end{equation}
 The corresponding field equations of the M$_{44}$ - model read as
 \begin{eqnarray}
 D_2F_{RR}+D_1F_R+KF&=&-2a^3\rho,\nonumber\\
   A_3F_{RRR}+A_2F_{RR}+A_1F_{R}+MF  &=&6a^2p,\label{3.30}\\
 \dot{\rho}+3H(\rho+p)&=&0.\nonumber
 \end{eqnarray}
 Here
 \begin{eqnarray} \label{3.31}
 D_2&=&-6\epsilon_1\dot{R}a^2\dot{a},\\ \label{3.32}
   D_1&=&6\epsilon_1a^2 \ddot{a}+a^3u_{\dot{a}}\dot{a},\\
   K&=&-a^3 \label{3.33}
 \end{eqnarray}
and
\begin{eqnarray}
  A_3&=&-6\epsilon_1\dot{R}^2a^2,\label{3.34}\\
A_2&=&-12\epsilon_1\dot{R}a\dot{a}-6\epsilon_1\ddot{R}a^2+a^3\dot{R}u_{\dot{a}},\label{3.35}\\
   A_1&=&12\epsilon_1\dot{a}^2+6\epsilon_1a \ddot{a}+3a^2\dot{a}u_{\dot{a}}+a^3\dot{u}_{\dot{a}}-a^3u_a,\label{3.36}\\
  M&=&-3a^2. \label{3.37}
 \end{eqnarray}
 If $u=0$ then we get the following equations of the  standard  $F(R_s)$ gravity (after $R=R_s$):
 \begin{eqnarray}
 6\dot{R}HF_{RR}-(R-6H^2)F_R+F&=&\rho,\label{3.38}\\
   -2\dot{R}^2F_{RRR}+[-4\dot{R}H-2\ddot{R}]F_{RR}+[-2H^2-4a^{-1} \ddot{a}+R]F_{R}-F &=&p,\label{3.39}\\
 \dot{\rho}+3H(\rho+p)&=&0.\label{3.40}
 \end{eqnarray}

 \textit{ii) The M$_{45}$ - model.}
The action of the M$_{45}$ - model looks like
 \begin{equation}\label{3.41}
 S_{45}=\int d^4 xe [F(T)+L_m],
 \end{equation}
 where $e={\rm det}\,(e_\mu^i)=\sqrt{-g}$ and the torsion scalar $T$ is defined as
 \begin{equation}\label{3.42}
 T=v+T_s=v+\epsilon_2{S_\rho}^{\mu\nu}\,{T^\rho}_{\mu\nu}.
 \end{equation}
 Here
 \begin{eqnarray}
 {T^\rho}_{\mu\nu} &\equiv &-e^\rho_i\left(\partial_\mu e^i_\nu
 -\partial_\nu e^i_\mu\right),\label{3.43}\\
 {K^{\mu\nu}}_\rho &\equiv &-\frac{1}{2}\left({T^{\mu\nu}}_\rho
 -{T^{\nu\mu}}_\rho-{T_\rho}^{\mu\nu}\right), \label{3.44}\\
 {S_\rho}^{\mu\nu} &\equiv &\frac{1}{2}\left({K^{\mu\nu}}_\rho
 +\delta^\mu_\rho {T^{\theta\nu}}_\theta-
 \delta^\nu_\rho {T^{\theta\mu}}_\theta\right).\label{3.45}
 \end{eqnarray}
 For a spatially flat FRW metric \eqref{2.18},  we have the torsion scalar  in the form
 \begin{equation}\label{3.46}
 T=v+T_s=v+6\epsilon_2H^2.
 \end{equation}
 The action \eqref{3.41} can be written as
 \begin{equation}\label{3.47}
 S_{45}=\int dt L_{45},
 \end{equation}
where
 the point-like Lagrangian reads
 \begin{equation}\label{3.48}
  L_{45}=a^3[F-(T-v)F_T+L_m]+6\epsilon_2F_Ta\dot{a}^2.
 \end{equation}
  So finally we get the following  equations of the M$_{45}$ - model:
  \begin{eqnarray}
 E_1F_T+KF&=&-2a^3\rho,\nonumber\\
   B_2F_{TT}+B_1F_{T}+MF  &=&6a^2p,\label{3.49}\\
 \dot{\rho}+3H(\rho+p)&=&0.\nonumber
 \end{eqnarray}
 Here
 \begin{eqnarray} \label{3.50}
  E_1&=&12\epsilon_2a \dot{a}^2+a^3v_{\dot{a}}\dot{a},\\
 K&=&-a^3  \label{3.51}
 \end{eqnarray}
and
\begin{eqnarray} \label{3.52}
  B_2&=&12\epsilon_2\dot{T}a \dot{a}+a^3\dot{T}v_{\dot{a}},\\
 B_1&=&24\epsilon_2\dot{a}^2+12\epsilon_2a     \ddot{a}+3a^2\dot{a}v_{\dot{a}}+a^3\dot{v}_{\dot{a}}-a^3v_a, \label{3.53} \\     M&=&-3a^2.       \label{3.54}
 \end{eqnarray}
 If we put $v=0$ then the M$_{45}$ - model reduces to the usual $F(T_s)$ gravity, where
 $T_s=6\epsilon_2H^2$. As is well-known the equations of $F(T_s)$ gravity are given by
 \begin{eqnarray}
 12H^2 F_T+F&=&\rho,\label{3.55}\\
 48H^2 F_{TT}\dot{H}-F_T\left(12H^2+4\dot{H}\right)-F
 &=&p,\label{3.56}\\
 \dot{\rho}+3H(\rho+p)&=&0,\label{3.57}
 \end{eqnarray}
 where we must put $T=T_s$. Finally we note that it is well-known that the standard
 $F(T_s)$ gravity is not local Lorentz invariant \cite{Sotiriou}.
  In this context, we have a very meager hope that the M$_{45}$ - model
  \eqref{3.41}  is free from such problems.

 \section{Cosmological solutions}
 In this section we investigate cosmological consequences of the $F(R,T)$ gravity. As example,  we want to find some exact cosmological solutions of the M$_{37}$ - gravity model.
 Since its equations are very complicated we here consider the simplest case when
 \begin{equation}\label{4.1}
 F(R,T)=\mu R+\nu T,
 \end{equation}
 where $\mu$ and $\nu$ are some constants. Then equations \eqref{3.9} take the form
  \begin{eqnarray}
 \mu D_1+\nu E_1+K(\nu T+\mu R)&=&-2a^3\rho,\nonumber\\
   \mu A_1+\nu B_1+M(\nu T+\mu R) &=&6a^2p,\label{4.2}\\
 \dot{\rho}+3H(\rho+p)&=&0,\nonumber
 \end{eqnarray}
 where
 \begin{eqnarray}
    D_1&=&6\epsilon_1a^2 \ddot{a}+a^3u_{\dot{a}}\dot{a},\label{4.3}\\
   E_1&=&12\epsilon_2a \dot{a}^2+a^3v_{\dot{a}}\dot{a},\label{4.4}\\
 K&=&-a^3,\label{4.5}\\
   A_1&=&12\epsilon_1\dot{a}^2+6\epsilon_1a \ddot{a}+3a^2\dot{a}u_{\dot{a}}+a^3\dot{u}_{\dot{a}}-a^3u_a,\label{4.6}\\
 B_1&=&24\epsilon_2\dot{a}^2+12\epsilon_2a \ddot{a}+3a^2\dot{a}v_{\dot{a}}+a^3\dot{v}_{\dot{a}}-a^3v_a, \label{4.7}\\
  M&=&-3a^2. \label{4.8}
 \end{eqnarray}

 We can rewrite this system as
 \begin{eqnarray}
 3\sigma H^2-0.5(\dot{a}z_{\dot{a}}-z)&=&\rho,\nonumber \\
-\sigma(2\dot{H}+3H^2)+0.5(\dot{a}z_{\dot{a}}-z)+\frac{1}{6}a(\dot{z}_{\dot{a}}-z_a)&=&p,\label{4.9}\\
 \dot{\rho}+3H(\rho+p)&=&0,\nonumber
 \end{eqnarray}
 where $z=\mu u+\nu v,\quad \sigma=\mu\epsilon_1-\nu\epsilon_2$. This system contents two independent
 equations for  five
  unknown functions ($a,\rho, p, u, v$). But in fact it contain 4 unknown functions ($H,\rho, p, z$).
   The corresponding  EoS parameter reads as
 \begin{equation}\label{4.10}
 \omega= \frac{p}{\rho}=-1+\frac{-2\sigma\dot{H}+\frac{1}{6}a(\dot{z}_{\dot{a}}-z_a)}{3\sigma H^2-0.5(\dot{a}z_{\dot{a}}-z)}.
 \end{equation}
 Let us find some simplest cosmological solutions of the system (\ref{4.9}). %\subsection{Case: $u=\alpha a^n, \quad v=\beta a^m$}
\subsection{Example 1} We start from the case $\sigma=0$. In this case the system \eqref{4.9} takes the form
\begin{eqnarray}
 -0.5(\dot{a}z_{\dot{a}}-z)&=&\rho,\nonumber \\
0.5(\dot{a}z_{\dot{a}}-z)+\frac{1}{6}a(\dot{z}_{\dot{a}}-z_a)&=&p,\label{4.11}\\
 \dot{\rho}+3H(\rho+p)&=&0.\nonumber
 \end{eqnarray}
At the same time  the EoS parameter becomes
 \begin{equation}\label{4.12}
 \omega= \frac{p}{\rho}=-1-\frac{a(\dot{z}_{\dot{a}}-z_a)}{3(\dot{a}z_{\dot{a}}-z)}.
 \end{equation}
 Now we assume that
  \begin{equation}\label{4.13}
 z=\kappa  a^l,
 \end{equation}
 where $\kappa$ and $l$ are some real constants.  Then
 \begin{equation}\label{4.14}
 \omega= -1-\frac{l}{3}.
 \end{equation}
 This result tells us that in this case our model can describes the accelerated expansion of the Universe for some values of $l$.
 \subsection{Example 2}
 Now consider the de Sitter case that is $H=H_0=const$ so that $a=a_0e^{H_0t}$.  Then the system \eqref{4.9} reads as
 \begin{eqnarray}\label{4.15}
 3\sigma H_0^2-0.5(\dot{a}z_{\dot{a}}-z)&=&\rho,\nonumber \\
-3\sigma H_0^2+0.5(\dot{a}z_{\dot{a}}-z)+\frac{1}{6}a(\dot{z}_{\dot{a}}-z_a)&=&p,\\
 \dot{\rho}+3H(\rho+p)&=&0.\nonumber
 \end{eqnarray}
 The  EoS parameter takes the form
 \begin{equation}\label{4.16}
 \omega= \frac{p}{\rho}=-1+\frac{a(\dot{z}_{\dot{a}}-z_a)}{18\sigma H^2_0-3(\dot{a}z_{\dot{a}}-z)}.
 \end{equation}
  If $z$ has the form \eqref{4.13} that is $z=\kappa e^{H_0lt}$ then we have
   \begin{equation}\label{4.17}
 \omega= \frac{p}{\rho}=-1-\frac{l\kappa}{18\sigma H^2_0e^{-H_0lt}+3\kappa}.
 \end{equation}
 If $H_0l>0$ then as $t\rightarrow\infty$ we get again
 \begin{equation}\label{4.18}
 \omega= -1-\frac{l}{3},
 \end{equation}
which is same as equation \eqref{4.14}. This last equation  tells us
that our model can describes the accelerated expansion of the
Universe e.g. for $l\geq 0$. Also it corresponds to the phantom case if
$l>0$. Finally we present other forms of the generalized Friedmann
equations in the system \eqref{4.9}. Let us rewrite these equations
in the standard form as
 \begin{eqnarray}
 3 H^2&=&\rho+\rho_z,\label{4.19} \\
-(2\dot{H}+3H^2)&=&p+p_z.\label{4.20}
 \end{eqnarray}
 Here
 \begin{eqnarray}
 \rho_z&=&0.5\sigma^{-1}(\dot{a}z_{\dot{a}}-z),\label{4.21} \\
p_z&=&-0.5\sigma^{-1}(\dot{a}z_{\dot{a}}-z)-\frac{1}{6\sigma}a(\dot{z}_{\dot{a}}-z_a)\label{4.22}
 \end{eqnarray}
 are the $z$ or $u-v$ contributions to the energy density and pressure, respectively.

\section{Conclusion}

In \cite{Buchdahl}, Buchdahl proposed  to replace the
Einstein-Hilbert scalar Lagrangian $R$  with a function of the
scalar curvature. The resulting theory is nowadays known as $F(R)$
gravity. Almost 40 years later, Bengochea and Ferraro proposed to
replace the TEGR that is the torsion scalar Lagrangian $T$ with a
function $F(T)$ of the torsion scalar, and studied its cosmological
consequences \cite{Bengochea}. This type of modified gravity is
nowadays called as $F(T)$ gravity theory. These two gravity theories
[that is  $F(R)$ and $F(T)$]  are, in some sense, alternative ways
to modify GR. From these results arises the natural question: how we
can construct some modified gravity theory which unifies $F(R)$ and
$F(T)$  theories?  Examples of such unified curvature-torsion
theories were proposed  in \cite{MK1}-\cite{MK1A}. Such type of
modified gravity theory is called the $F(R,T)$ gravity. In this
$F(R,T)$ gravity, the curvature scalar $R$ and the torsion scalar
$T$ play the same role and are dynamical quantities. In this paper,
we have shown that the $F(R,T)$ gravity can be derived from the
geometrical point of view. In particular, we have proposed a new
method to construct particular models of $F(R,T)$ gravity. As an
  example we have considered the M$_{43}$ - model, deriving its action
   in terms of the curvature and torsion scalars. Then in detail we
    have studied the M$_{37}$ - model and presented its action,
     Lagrangian and equations of motion for the FRW metric case.
      Finally we have shown that the last model can describes the
       accelerated expansion of the Universe.

Concluding, we would like to note that in the paper, we  present a  special class of extended gravity
models depending on  arbitrary function $F(R,T)$, where $R$ is the
Ricci scalar and $T$ the scalar torsion. While in the traditional
Eistein-Cartan theory, the role of the torsion depends on the non
trivial source associated with spin matter density, in our $F(R,T)$
gravity models,  the torsion can propagate without the presence of
spin matter density. In fact this is a crucial point, otherwise the
additional scalar torsion degree of freedom are not different from
the additional metric gravitational  degree of freedom present in
extended $F(R)$ models.   Finally we would like to note that all
results of this paper are new and different than results of our
previous papers  \cite{MK1}-\cite{MK1A} on the subject.


\begin{thebibliography}{99}

\bibitem{j1} M. U. Farooq, M. Jamil, U. Debnath, Astrophys. Space
Sci. \textbf{334}, 243 (2011).

\bibitem{j2} M. Jamil, D. Momeni, K. Bamba, R. Myrzakulov, Int. J. Mod. Phys. D \textbf{21} (2012)
1250065;\\ M. Jamil, I. Hussain, M. U. Farooq, Astrophys. Space Sci.
\textbf{335}, 339 (2011);\\ M. Jamil, I. Hussain, Int. J. Theor.
Phys. \textbf{50}, 465 (2011).

\bibitem{j3} A. Pasqua, A. Khodam-Mohammadi, M. Jamil, R.
Myrzakulov,  Astrophys. Space Sci. \textbf{340,} 199 (2012).

\bibitem{j4} M. U. Farooq, M. A. Rashid, M. Jamil,  Int. J. Theor. Phys. \textbf{49,} 2278
(2010);\\ K. Karami, M. S. Khaledian, M. Jamil, Phys. Scr.\textbf{
83,} 025901 (2011);\\ M. Jamil, F. M. Mahomed, D. Momeni, Phys.
Lett. B \textbf{702,} 315
 (2011);\\ M. Jamil, A. Sheykhi, Int. J. Theor. Phys.\textbf{ 50,} 625 (2011).

\bibitem{j5} R. Myrzakulov, arXiv:1011.4337;  \\ R. Myrzakulov, arXiv:1103.5918; \\ M. Jamil, D. Momeni, N. S. Serikbayev, R. Myrzakulov, Astrophys. Space Sci.  \textbf{339,} 37
(2012);\\ M. Jamil, Y. Myrzakulov, O. Razina, R. Myrzakulov,
Astrophys. Space Sci.  \textbf{336,} 315  (2011);
\\ O.V. Razina, Y.M. Myrzakulov, N.S. Serikbayev, G. Nugmanova, R. Myrzakulov,
Cent. Eur. J. Phys.,   \textbf{10}, N1, 47-50  (2012); \\ I. Kulnazarov, K.  Yerzhanov, O. Razina, Sh. Myrzakul, P. Tsyba, R. Myrzakulov,
Eur. Phys. J. C,   \textbf{71}, 1698  (2011).

\bibitem{j6} S. Chakraborty, U. Debnath, M. Jamil, R. Myrzakulov,  Int. J. Theor. Phys. \textbf{ 51,} 2246
(2012);\\ M.U.  Farooq, M. A. Rashid, M. Jamil, Int. J. Theor. Phys.
\textbf{49,} 2334 (2010);\\ M. Jamil, M.U. Farooq,  JCAP
\textbf{03,} 001 (2010);\\ M. Jamil,  Int. J. Theor. Phys.
\textbf{49,} 62
(2010);\\ M. Jamil, M.A. Rashid, Eur. Phys. J. C \textbf{58,} 111 (2008);\\
M. Jamil, M.A. Rashid, Eur. Phys. J. C \textbf{60,} 141 (2009).

\bibitem{r1}
E.~J.~Copeland, M.~Sami and S.~Tsujikawa,
 Int.\ J.\ Mod.\ Phys.\  D {\bf 15}, 1753 (2006)
 [hep-th/0603057];\\
J.~Frieman, M.~Turner and D.~Huterer,
 Ann.\ Rev.\ Astron.\ Astrophys.\  {\bf 46}, 385 (2008)
 [arXiv:0803.0982];\\
S.~Tsujikawa, arXiv:1004.1493 [astro-ph.CO];\\
M.~Li, X.~D.~Li, S.~Wang and Y.~Wang,
 Commun.\ Theor.\ Phys.\  {\bf 56}, 525 (2011) [arXiv:1103.5870];\\
Y.~F.~Cai, E.~N.~Saridakis, M.~R.~Setare and J.~Q.~Xia,
 Phys.\ Rept.\  {\bf 493}, 1 (2010) [arXiv:0909.2776].

\bibitem{r2}
A.~De Felice and S.~Tsujikawa,
 Living Rev.\ Rel.\  {\bf 13}, 3 (2010) [arXiv:1002.4928];\\
T.~Clifton, P.~G.~Ferreira, A.~Padilla and C.~Skordis,
 arXiv:1106.2476 [astro-ph.CO];\\
T.~P.~Sotiriou and V.~Faraoni,
 Rev.\ Mod.\ Phys.\  {\bf 82}, 451 (2010) [arXiv:0805.1726];\\
S.~Tsujikawa,
 Lect.\ Notes Phys.\  {\bf 800}, 99 (2010) [arXiv:1101.0191];\\
S.~Capozziello, M.~De Laurentis and V.~Faraoni,
 arXiv:0909.4672 [gr-qc];\\
R.~Durrer and R.~Maartens, arXiv:0811.4132 [astro-ph];\\
S.~Nojiri and S.~D.~Odintsov,
 Int.\ J.\ Geom.\ Meth.\ Mod.\ Phys.\  {\bf 4}, 115 (2007)
 [hep-th/0601213];\\
S.~Nojiri and S.~D.~Odintsov, arXiv:1011.0544 [gr-qc];\\
M.R. Setare, M. Jamil, Gen. Relativ. Gravit.  \textbf{43}, 293 (2011);\\
M. Jamil, E. N. Saridakis, M. R. Setare, JCAP \textbf{1011}, 032
(2010).


\bibitem{Ba} K. Bamba, S. Capozziello, S. Nojiri, S. D. Odintsov,  arXiv:1205.3421.\\
S.~Capozziello, M.~De Laurentis and V.~Faraoni,
 arXiv:0909.4672 [gr-qc];\\
S.~Nojiri and S.~D.~Odintsov,
 Int.\ J.\ Geom.\ Meth.\ Mod.\ Phys.\  {\bf 4}, 115 (2007)
 [hep-th/0601213];\\
S.~Nojiri and S.~D.~Odintsov, arXiv:1011.0544 [gr-qc];\\
M. Jamil, D. Momeni, R. Myrzakulov, Eur. Phys. J. C  \textbf{72,}
2122 (2012);\\ M. Jamil, D. Momeni, R. Myrzakulov, Eur. Phys. J. C
\textbf{72}, 2075 (2012);\\  M. Jamil, D. Momeni, R. Myrzakulov,
Eur. Phys. J. C
 \textbf{72,} 1959 (2012);\\M. Jamil, D. Momeni, R. Myrzakulov, Eur. Phys. J.
 C \textbf{72,} 2137 (2012).



\bibitem{MK1} R. Myrzakulov,  General Relativity and Gravitation, {\bf 44},  (2012)  [DOI:10.1007/s10714-012-1439-z], [arXiv:1008.4486]
\bibitem{MK1A} R. Myrzakulov.  arXiv:1205.5266; \\ 
R. Myrzakulov, Entropy, \textbf{14}, N9, 1627-1651 (2012); \\
M. Sharif, S. Rani, R. Myrzakulov, arXiv:1210.2714; \\
S. Chattopadhyay, arXiv:1208.3896.

\bibitem{T}
F. Muller-Hoissen. Phys. Lett. A,  {\bf 92}, N9, 433-434 (1982);\\
A. Dimakis, F. Muller-Hoissen. Phys. Lett. A,  {\bf 92}, N9, 431-432 (1982);\\
F. Muller-Hoissen, J. Nitsch. Phys. Rev. D,  {\bf 28}, N4, 718-728 (1983);\\
A. Dimakis, F. Muller-Hoissen. J. Math. Phys.,  {\bf 26}, N5, 1040-1048 (1985);\\
F. Muller-Hoissen, J. Nitsch. Gen. Rel. Grav.,  {\bf 17}, N8, 747-760 (1985);\\
H. Goenner, F. Muller-Hoissen. Class. Quantum Grav.,  {\bf 1}, 651-672 (1984);\\
F. Muller-Hoissen. Ann. Inst. Henri Poincare,  {\bf 40}, N1, 21-34 (1984);\\
S.~Capozziello and R.~de Ritis,
 Class.\ Quant.\ Grav.\  {\bf 11}, 107 (1994);\\
 M. Tsamparlis,
 Phys. Lett. A,   {\bf 75}, N1,2, 27-28  (1979);\\
 D.P. Mason, M. Tsamparlis,
 General Relativity and Gravitation,   {\bf 13}, N2, 123-134  (1981);\\
 A.V. Minkevich, A.S. Garkun , [gr-qc/9805007];\\
 S. Capozziello, G. Lambiase, C. Stornaiolo,
 Ann. Phys. (Leipzig),   {\bf 10}, N8, 713-727  (2001).


\bibitem{r5}
G.~R.~Bengochea and R.~Ferraro,
 Phys.\ Rev.\  D {\bf 79}, 124019 (2009) [arXiv:0812.1205].

\bibitem{r6a} R.~Myrzakulov,
 Eur.\ Phys.\ J.\  C {\bf 71}, 1752 (2011) [arXiv:1006.1120]
\bibitem{r6b} K.~K.~Yerzhanov, S.~R.~Myrzakul, I.~I.~Kulnazarov and R.~Myrzakulov,
 [arXiv:1006.3879]
 \bibitem{MK0} R. Myrzakulov, Advances in High Energy Physics, {\bf 2012}, 868203 (2012) [arXiv:1204.1093]
 \bibitem{MK00} K.R. Yesmakhanova, N. A. Myrzakulov, K. K. Yerzhanov, G.N. Nugmanova, N.S. Serikbayaev, R. Myrzakulov, [arXiv:1201.4360]
 \bibitem{C1} S. Capozziello, M. De Laurentis, L. Fabbri, S. Vignolo,  Eur. Phys. J. C, {\bf 72}, 1908-1918, 2012 [arXiv:1202.3573]
 \bibitem{Fabbri} L. Fabbri, S. Vignolo,  Int. J. Theor. Phys. {\bf 51}, 3186-3207 (2012) [arXiv:1201.5498]
 \bibitem{Ao} X.-C. Ao, X.-Z. Li. JCAP, {\bf  1110}, 039 (2011).  [arXiv:1111.1801]
 \bibitem{Qi}  G.-Y. Qi.   [arXiv:1110.3449]
 \bibitem{Poplawski}  N. Poplawski.   [arXiv:1203.0294] 
\bibitem{MK2} A.J. Lopez-Revelles, R.  Myrzakulov, D. Saez-Gomez, Physical Review D, {\bf 85}, N10, 103521 (2012).
\bibitem{MK3} K. Bamba, R. Myrzakulov, S. Nojiri, S. D. Odintsov,  Physical Review D, {\bf 85}, N10, 104036 (2012).

\bibitem{MK4} M. Duncan, R.
Myrzakulov, D. Singleton.  Phys. Lett. B, {\bf 703},  N4, 516-518 (2011).

\bibitem{MK5} R. Myrzakulov, D. Saez-Gomez, A. Tureanu. General Relativity and Gravitation, {\bf 43}, N6, 1671-1684  (2011)

\bibitem{MK6} V. Dzhunushaliev, V.
Folomeev, R.
Myrzakulov, D. Singleton. Physical Review D,
{\bf 82},  045032 (2010)
\bibitem{MK6A} I. Brevik, R. Myrzakulov, S. Nojiri,  S. D. Odintsov. Physical Review D,
{\bf 86}, N6,  063007 (2012)
\bibitem{MK7} E. Elizalde, R. Myrzakulov, V.V. Obukhov, D. Saez-Gomez. Classical and Quantum Gravity, {\bf 27}, N8, 085001-12 (2010)
  \bibitem{Buchdahl}  H.A. Buchdahl.  Mon. Not. R. Astron. Soc., {\bf 150}, 1 (1970).

  \bibitem{Bengochea} G.R. Bengochea, R. Ferraro. Phys. Rev. D,  {\bf 79}, 124019  (2009).
  \bibitem{Harko} Harko T., Lobo F.S.N., Nojiri S., Odintsov S.D.  Phys. Rev. D,  {\bf 84}, 024020  (2011).
  \bibitem{Sotiriou} Sotiriou T.P., Li B., Barrow J.D.  Phys. Rev. D,  {\bf 83}, 104030  (2011).

\end{thebibliography}
\end{document}